\documentclass{article}
\usepackage{epsfig}
\title{Unambiguous discrimination of special sets of multipartite 
states using local measurements and classical communication}
\author{Jihane Mimih and Mark Hillery \\ Department of Physics and
Astronomy \\ Hunter College of CUNY \\ 695 Park Avenue \\ New
York, NY 10021}
\begin{document}
\maketitle
\begin{abstract}
We initially consider a quantum system consisting of two qubits, which
can be in one of two nonorthogonal 
states, $|\Psi_{0}\rangle$ and$|\Psi_{1}\rangle$. We distribute
the qubits to two parties, Alice and Bob. They each measure
their qubits and then compare their measurement results to determine 
which state they were sent.  This procedure is error-free, which implies
that it must sometimes fail.  In addition, no quantum memory is required;
it is not necessary for one of the qubits to be stored until the result
of the measurement on the other is known.  
We consider the cases in which, should failure
occur, both parties receive a failure signal or only one does.  In the
latter case, if the two states share the same Schmidt basis, the states can 
be discriminated with the same failure probability that would be obtained if
the qubits were measured together.  This scheme is sufficiently simple that
it can be generalized to multipartite qubit, and qudit, states.  Applications
to quantum secret sharing are discussed.  Finally, we present an optical 
scheme to experimentally realize the protocol in the case of two qubits.
\end{abstract}

\section{Introduction}
Suppose we have two qubits prepared in one of two quantum states,
$|\Psi_{0}\rangle$ or $|\Psi_{1}\rangle$. We now give one qubit to Alice and 
one qubit to Bob. Both parties know that the state is either $|\Psi_{0}\rangle$
or $|\Psi_{1}\rangle$, and their task is to perform local measurements on
their qubits and communicate through a classical channel to determine
the state they have been given. Alice and Bob can perfectly distinguish
between the states using local operations and classical
communication only if the states are orthogonal \cite{walgate}. When
$|\Psi_{0}\rangle$ and$|\Psi_{1}\rangle$ are not orthogonal, Alice
and Bob can use two different strategies to distinguish between
the states.

The first one is the minimum error state discrimination approach.
In this case, after Alice and Bob measure their qubits, they have to give 
a conclusive answer about the state; they are not allowed to  
give ``don't know'' as an answer. However, since the states are 
not orthogonal, the price that the two parties must pay for giving
a definite answer is the chance that they will make a mistake and 
incorrectly identify the state. The minimum probability of making 
a wrong guess, when 
each state is equally likely, is \cite{virmani}
\begin{equation}
p_{E}=\frac{1}{2}(1-\sqrt{1-|\langle\Psi_{0}|\Psi_{1}\rangle|^{2}})
\end{equation}

An alternative approach to the state discrimination problem,
is the unambiguous state discrimination method. In this case, some 
measurement outcomes are allowed to be inconclusive; that is Alice and Bob 
might fail to identify the state, but if they succeed they will not make 
an error. If each state is equally likely and both qubits
are measured together, then the optimal probability to successfully
and unambiguously distinguish the states is \cite{ivanovic,dieks,peres}
 \begin{equation}
p_{idp}=1-|\langle\Psi_{0}|\Psi_{1}\rangle| .
\end{equation}

The probability of getting an inconclusive result, which provides no
information about the state, is $1-p_{idp}$.  This success probability can
also be achieved if the qubits are measured separately, one by Alice and 
one by Bob, and they are allowed to communicate through a classical 
channel \cite{virmani,chen}.  In this procedure, Alice makes a projective
measurement on her qubit that gives her no information about the state,
and she then communicates the result of her measurement to Bob.  Based on
this information, Bob is able to make a measurement on his qubit that
allows him to decide, with a success probability of $p_{idp}$, what the
initial state was.

If one wants to use this procedure as part of a quantum communication
scheme, in particular for secret sharing, there are difficulties. In a
secret sharing scheme, Alice and Bob are sent parts of a message or key
by a third party, Charlie,
and these parts have to be combined in order for the message or key to
be revealed \cite{hillery1,karlsson,cleve}.  The first problem, then, is
that if the parts are to be combined at a time significantly later than
when they were sent, quantum memory is required, i.e.\ the qubits have to
be protected against decoherence for a long time.  If one attempts to 
surmount this difficulty by having the parties measure their qubits
immediately upon receiving them, one is faced with the problem that the
information gain is asymmetric.  Alice learns nothing about the key, and
Bob learns everything. The only way this could be useful is if Alice and
Bob are in the same location and are to use the key immediately.  If they
are in separate locations and will be using the key later, another procedure
is required.

In a previous paper we discussed such a scheme \cite{hillery2}.  In it, 
both parties measure their qubit immediately upon receiving it, each 
obtaining a result of either $0$ or $1$. There are four sets of results: 
$\{0,0\}$, $\{1,1\}$,$\{0,1\}$, and $\{1,0\}$.  The result $\{0,0\}$
corresponds to $|\Psi_{0}\rangle$, the result $\{1,1\}$ corresponds to
$|\Psi_{1}\rangle$, and the results $\{0,1\}$, and $\{1,0\}$ correspond
to failure.  It was shown that in the case that the two states have the
same Schmidt basis, the probability of successfully identifying the state
is given by $p_{idp}$.  This procedure can be used in a secret-sharing
scheme; the set of measurement results obtained by Alice and Bob, which is 
classical information, can be stored indefinitely and compared at a later 
time to reveal the key.

This scheme does, however, have a drawback.  The key bits for which the 
measurement failed, and which, therefore, must be discarded, are only
identified after Alice and Bob have compared their bit strings.  It would
be much better if the bits that must be discarded could be identified
immediately.  The previous procedure requires that Alice and Bob get
together and then tell Charlie which bits are good and which are not.  He
can then send them a message.  A procedure in which the failed bits are
immediately identified, allows Charlie to send Alice and Bob the 
two-qubit states from which the key bits can be extracted, discard the 
failed bits, and then immediately send them the message.  At some later 
time, Alice and Bob can get together, combine their bit strings to get
the key, and then read the message, without further input from Charlie.
This latter scheme is much more flexible.

This can be accomplished by adding a third measurement result for one
or both of the parties. If this added result is obtained, the measurement
has failed to distinguish the states. In this paper, we will examine 
both the case in which both parties have three measurement 
outcomes, $0$, $1$, or $f$ for failure to distinguish, or only one does. 
In the latter case, the remaining party has only the outcomes $0$ and $1$.
We shall first examine the case in which both Alice and Bob receive a 
failure indication when the measurement fails.  We shall find that this
kind of scheme is impossible for two-qubit states if both states are to
be detected with a nonzero probability.  We shall then show that a procedure 
in which only one of the parties receives a failure signal is possible, 
and construct the necessary POVM.  In addition, we shall show how this 
procedure can be implemented optically.  The qubits are the polarization
states of photons.  Two photon states are created and one photon each
is sent to Alice and Bob.  Using linear optics they can perform the
necessary measurements and identify, with a certain probability, which of 
two possible two-photon states was sent.  Finally, we shall show 
how the procedure 
in which only one party receives a failure signal can be generalized to
$N$ parties, and to qudits rather than qubits.  The case of $N$ parties
discriminating among three $N$-qutrit states is discussed in detail.

\section{Failure indication received by both parties}
As discussed in the Introduction, we  shall first assume that the measurements
that Alice and Bob make have three possible outcomes, $0$, $1$, and $f$,
which denotes failure to distinguish . The POVM operators that characterize
the measurements are $\{ A_{0},A_{1}, A_{f}\}$ for Alice and 
$\{B_{0},B_{1},B_{f}\}$ for Bob.  These operators satisfy
\begin{equation}
\label{ident1} I_{A}=
\sum_{j=0,1,f}A_{j}^{\dagger}A_{j}\hspace{1cm}
I_{B}=\sum_{j=0,1,f} B_{j}^{\dagger}B_{j} ,
\end{equation}
where $I_{A}$ is the identity on $\mathcal{H}_{A}$, the Hilbert
space of Alice's qubit, and $I_{B}$ is the identity on
$\mathcal{H}_{B}$, the space of Bob's qubit.
We suppose that measurement results $\{ 0,0\}$ (Alice obtains $0$ and Bob
also obtains $0$) and $\{ 1,1\}$ correspond to
$|\Psi_{1}\rangle$, and $\{ 0,1\}$ and $\{ 1,0\}$ correspond to
$|\Psi_{0}\rangle$.  The reason for this choice is that we do not want Alice
or Bob to be able to tell from only the result of their measurement which
state was sent.  For example, if Alice always measured $0$ when 
$|\Psi_{0}\rangle$ was sent and $1$ when $|\Psi_{1}\rangle$ was sent, then 
she would have no need of any information from Bob to determine the
identity of the state. Consequently, for each state, Alice and Bob must have 
the possibility of receiving either a $0$ or a $1$.  The correspondence between
states and measurement results is one of only two choices that satisfies this
condition (the other simply switches the measurement results corresponding
to $|\Psi_{0}\rangle$ and $|\Psi_{1}\rangle$ ).
The condition that no errors are allowed requires that
\begin{eqnarray}
\label{cond} 
A_{0}B_{0}|\Psi_{0}\rangle = 0 \hspace{1cm}
A_{1}B_{1}|\Psi_{0}\rangle = 0 \\
A_{0}B_{1}|\Psi_{1}\rangle = 0
\hspace{1cm} A_{1}B_{0}|\Psi_{1}\rangle = 0 ,
\end{eqnarray}
and the condition that, if the measurement fails, then both Alice and
Bob find the result $f$, is
\begin{equation}
\label{simfail}
A_{f}B_{j}|\Psi_{k}\rangle = 0 \hspace{1cm}
A_{j}B_{f}|\Psi_{k}\rangle = 0 ,
\end{equation}
where $j=0,1$ and $k=0,1$

Expressing $|\Psi_{0}\rangle$ and $|\Psi_{1}\rangle$ in their Schmidt bases
we have
\begin{eqnarray}
|\Psi_{0}\rangle = \sum_{l=0}^{1}\sqrt{\lambda_{0l}}|u_{Al}\rangle
\otimes |u_{Bl}\rangle \\
|\Psi_{1}\rangle = \sum_{l=0}^{1}\sqrt{\lambda_{1l}}|v_{Al}\rangle \otimes
|v_{Bl}\rangle ,
\end{eqnarray}
where $\{ u_{A0}, u_{A1}\}$ and $\{ v_{A0}, v_{A1}\}$ are orthonormal bases 
for Alice's space, and $\{ u_{B0}, u_{B1}\}$ and $\{ v_{B0}, v_{B1}\}$ are 
orthonormal bases for Bob's space.  The coefficients $\lambda_{0l}$ and  
$\lambda_{1l}$ where $l=0,1$, are the eigenvalues of the reduced density 
matrices corresponding to $|\Psi_{0}\rangle$ and $|\Psi_{1}\rangle$, 
respectively.  Substituting this representation into the conditions in the
previous paragraph, we find, first, that the condition $A_{f}B_{j}|\Psi_{0}
\rangle = 0$ implies that
\begin{equation}
\sqrt{\lambda_{00}}A_{f} |u_{A0}\rangle \otimes B_{j}|u_{B0}\rangle =
-\sqrt{\lambda_{01}}A_{f} |u_{A1}\rangle \otimes B_{j}|u_{B1}\rangle .
\end{equation}
This is only possible if $ A_{f} |u_{A0}\rangle$ is parallel to 
$A_{f} |u_{A1}\rangle $ and if $B_{j} |u_{B0}\rangle$ is parallel to 
$B_{j} |u_{B1}\rangle$. Then, we have, for some vectors $|\eta_{Af}\rangle$ 
and $|\eta_{Bj}\rangle$, that
\begin{eqnarray}
\label{con1} 
A_{f}|u_{A0}\rangle = c_{0f}|\eta_{Af}\rangle &
B_{j}|u_{B0}\rangle = d_{j0}|\eta_{Bj}\rangle \nonumber \\
A_{f}|u_{A1}\rangle = c_{1f}|\eta_{Af}\rangle &
B_{j}|u_{B1}\rangle = d_{j1}|\eta_{Bj}\rangle ,
\end{eqnarray}
where $c_{lf}$ and $d_{jl}$ are constants, and $\|\eta_{Af}\|=1$
and $ \|\eta_{Bj}\| =1$. We can then express $A_{f}$ as
\begin{eqnarray}
A_{f} & =& |\eta_{Af}\rangle( c_{0f} \langle u_{A0}| + c_{1f} \langle
u_{A1}|) \nonumber \\
 & = & x_{f}|\eta_{Af}\rangle\langle r_{f}| ,
\end{eqnarray}
where
\begin{equation}
|r_{f}\rangle =\frac{1}{(|c_{0f}|^{2}+|c_{1f}|^{2})^{1/2}}(c_{0f}^{\ast}
|u_{A0}\rangle +c_{1f}^{\ast}|u_{A1}\rangle ) .
\end{equation}  
Similarly, we find that for $j=0,1,f$
\begin{equation}
\label{con2}
A_{j}=x_{j}|\eta_{Aj}\rangle \langle r_{j}|\hspace{1cm} 
B_{j}=y_{j}|\eta_{Bj}\rangle \langle s_{j}| ,
\end{equation}
where $|\eta_{Aj}\rangle$, $|\eta_{Bj}\rangle$, $|r_{j}\rangle$, and
$|s_{j}\rangle$ are unit vectors, and the constants $x_{j}$ and $y_{j}$
are yet to be determined.

We can substitute the above expressions for the POVM operators into the
conditions for no errors and for simultaneous failure results, Eqs.\ 
(\ref{cond}) and (\ref{simfail}).  The equations containing 
$|\Psi_{0}\rangle$ are
\begin{eqnarray}
\sqrt{\lambda_{00}}\langle r_{0}|u_{A0}\rangle \langle
s_{0}|u_{B0}\rangle+\sqrt{\lambda_{01}}\langle r_{0}|u_{A1}\rangle
\langle s_{0}|u_{B1}\rangle=0 \nonumber \\
\sqrt{\lambda_{00}}\langle r_{1}|u_{A0}\rangle \langle
s_{1}|u_{B0}\rangle+\sqrt{\lambda_{01}}\langle r_{1}|u_{A1}\rangle
\langle s_{1}|u_{B1}\rangle=0 \nonumber \\
\sqrt{\lambda_{00}}\langle r_{f}|u_{A0}\rangle \langle
s_{j}|u_{B0}\rangle+\sqrt{\lambda_{01}}\langle r_{f}|u_{A1}\rangle
\langle s_{j}|u_{B1}\rangle=0 \nonumber \\
\sqrt{\lambda_{00}}\langle r_{j}|u_{A0}\rangle \langle
s_{f}|u_{B0}\rangle+\sqrt{\lambda_{01}}\langle r_{j}|u_{A1}\rangle
\langle s_{f}|u_{B1}\rangle=0 .
\end{eqnarray}
Defining the matrix
\begin{equation}
 M^{(0)}=\left(\begin{array}{cc}\sqrt{\lambda_{00}}  & 0 \\ 
0 & \sqrt{\lambda_{01}} \end{array} \right)
\end{equation}
and the vectors
\begin{equation}
\overline{r}_{j}^{*}=\left(\begin{array}{c}\langle r_{j}|u_{A0}\rangle  \\
\langle r_{j}|u_{A1}\rangle\end{array} \right)
\hspace{1cm}
\overline{s}_{j}=\left(\begin{array}{c}\langle s_{j}|u_{B0}\rangle  \\
\langle s_{j}|u_{B1}\rangle \end{array} \right) ,
\end{equation}
we can express the  above equations as 
\begin{eqnarray}
\label{M0}
\overline{r}_{0}^{*} \cdot M^{(0)}\overline{s}_{0}=0 \hspace{1cm} &
\overline{r}_{f}^{*} \cdot M^{(0)}\overline{s}_{j}=0 \nonumber \\
\overline{r}_{1}^{*} \cdot M^{(0)}\overline{s}_{1}=0 \hspace{1cm} &
\overline{r}_{j}^{*} \cdot M^{(0)}\overline{s}_{f}=0 .
\end{eqnarray}

It is straightforward to show that if both $\lambda_{00}$ and $\lambda_{01}$ 
are not zero, and if $\overline{w}^{\ast}\cdot M^{(0)}\overline{x} =0$ and 
$\overline{w}^{\ast}\cdot M^{(0)}\overline{y}=0$, for $\overline{w}\neq 0$,
then $\overline{x}$ is
a multiple of $\overline{y}$.  Applying this to Eqs.\ (\ref{M0}), we see
that $\overline{s}_{1}$ is a multiple of $\overline{s}_{f}$  and that 
$\overline{s}_{0}$ is also a multiple of $\overline{s}_{f}$ . The fact that 
the three vectors, $\overline{s}_{0}$, $\overline{s}_{1}$ and 
$\overline{s}_{f}$, are parallel violates the condition
$I_{B}=\sum_{j=0,1,f} y_{j}|s_{j}\rangle \langle s_{j}|$.  If we attempt to
circumvent this by choosing either $|r_{0}\rangle$ or $|r_{1}\rangle$ equal
to zero, we still find that $\overline{s}_{0}$, $\overline{s}_{1}$ and 
$\overline{s}_{f}$ are parallel.   

The cases in which either $\lambda_{00}$ or $\lambda_{01}$ are zero also 
need to be examined, but the conclusion is the same; it is not possible to
construct a POVM that satisfies Eqs.\ (\ref{cond}) and (\ref{simfail})
and for which both $|\Psi_{0}\rangle$ and $|\Psi_{1}\rangle$ have
a nonzero probability of being detected.
There are simply too many restrictions on the POVM elements and they cannot 
all be satisfied.  Therefore, we cannot construct a POVM that is error-free, 
and for which Alice and Bob receive simultaneous failure signals, when the
procedure fails.  It should be noted, as shown in \cite{hillery2}, that 
if qutrits are used instead of qubits, an error-free POVM with simultaneous 
failure signals is possible.

\section{ Failure signal received by one party}
In light of what we have just learned it makes sense to now consider the 
situation in which only one party receives a failure indication when the 
measurement fails.  In particular, both parties will have the possibility
of receiving a failure signal, and if either one of them does (even if the
other does not), then the procedure has failed.  No assumption is made
about which party will receive a failure signal. 
We shall also consider a special case, that in which
$|\Psi_{0}\rangle$ and $|\Psi_{1}\rangle$ have the same Schmidt bases and 
are given by
\begin{eqnarray}
|\Psi_{0}\rangle=\cos\theta_{0} |00\rangle+\sin\theta_{0}|11\rangle 
\nonumber \\
|\Psi_{1}\rangle=\cos\theta_{1} |00\rangle+\sin\theta_{1}
|11\rangle .
\end{eqnarray}
The conditions that no errors are allowed are the same as before
\begin{eqnarray}
A_{0}B_{0}|\Psi_{0}\rangle=0 \hspace{1cm}
A_{1}B_{1}|\Psi_{0}\rangle=0 \nonumber\\
A_{0}B_{1}|\Psi_{1}\rangle=0 \hspace{1cm}
A_{1}B_{0}|\Psi_{1}\rangle=0 .
\end{eqnarray}
These conditions imply, as before, that for $j=0,1$
\begin{eqnarray}
A_{j}=x_{j}|\eta_{Aj}\rangle \langle r_{j}|\hspace{1cm}
B_{j}=y_{j}|\eta_{Bj}\rangle \langle s_{j}| ,  \nonumber\\
\end{eqnarray}
and we shall express the vectors $|r_{j}\rangle$ and $|s_{j}\rangle$ in the
basis $\{ |0\rangle ,|1\rangle \}$ as
\begin{eqnarray}
|r_{0}\rangle=a_{0}|0\rangle+a_{1}|1\rangle \hspace{1cm}
|s_{0}\rangle=c_{0}|0\rangle+c_{1}|1\rangle \nonumber\\
|r_{1}\rangle=b_{0}|0\rangle+b_{1}|1\rangle \hspace{1cm}
|s_{1}\rangle=d_{0}|0\rangle+d_{1}|1\rangle
\end{eqnarray}
The no-error conditions can now be expressed as
\begin{eqnarray}
\label{rsnoerror}
(\langle r_{0}|\langle s_{0}|)|\Psi_{0}\rangle=0 \hspace{1cm}
(\langle r_{1}|\langle s_{1}|)|\Psi_{0}\rangle=0 \nonumber\\
(\langle r_{0}|\langle s_{1}|)|\Psi_{1}\rangle=0 \hspace{1cm} (\langle
r_{1}|\langle s_{0}|)|\Psi_{1}\rangle=0 .
\end{eqnarray}
Defining the ratios
\begin{eqnarray}
z_{0}=\frac{a_{1}}{a_{0}} & z_{1}=\frac{b_{1}}{b_{0}}  \\
z_{2}=\frac{c_{1}}{c_{0}} & z_{3}=\frac{d_{1}} {d_{0}} ,
\end{eqnarray}
Equations (\ref{rsnoerror}) become
\begin{eqnarray}
1+z_{0}^{\ast}z_{2}^{\ast} \tan\theta_{0}=0 \hspace{1cm}
1+z_{1}^{\ast}z_{3}^{\ast} \tan\theta_{0}=0 \nonumber \\
1+z_{0}^{\ast}z_{3}^{\ast} \tan\theta_{1}=0 \hspace{1cm}
1+z_{1}^{\ast}z_{2}^{\ast} \tan\theta_{1}=0
\end{eqnarray}

A necessary condition for these equations to have a solution is that
$\tan\theta_{0}=\pm\tan\theta_{1}$. We are not interested in the case where
$\tan\theta_{0}= \tan\theta_{1}$, since this implies that our states are 
identical. We wish to examine the case where $\tan\theta_{0}
= - \tan\theta_{1}$, which implies that $\theta_{1}=-\theta_{0}$. Hence, 
our two states can be expressed as
\begin{eqnarray}
\label{states}
|\Psi_{0}\rangle=\cos\theta_{0} |00\rangle+\sin\theta_{0}|11\rangle 
\nonumber \\
|\Psi_{1}\rangle=\cos\theta_{0} |00\rangle-\sin\theta_{0}|11\rangle
\end{eqnarray}
In this case, we find
\begin{equation}
z_{2}=-\frac{1}{z_{0}} \cot\theta_{0} \hspace{1cm}
z_{3}=\frac{1}{z_{0}} \cot \theta_{0} \hspace{1cm} z_{1}=-z_{0} .
\end{equation}
We can now express the vectors $|r_{j}\rangle$ and $|s_{j}\rangle$ as
\begin{eqnarray}
|r_{0}\rangle=
\frac{1}{\sqrt{1+|z_{0}|^{2}}}(|0\rangle+z_{0}|1\rangle)
\nonumber \\
|r_{1}\rangle=
\frac{1}{\sqrt{1+|z_{0}|^{2}}}(|0\rangle-z_{0}|1\rangle)
\nonumber\\
|s_{0}\rangle=
\sqrt{\frac{|z_{0}|^{2}}{|z_{0}|^{2}+{\cot\theta_{0}}^{2}}}(|0\rangle-\frac{\cot\theta_{0}}{z_{0}}|1\rangle)
\nonumber\\
|s_{1}\rangle=
\sqrt{\frac{|z_{0}|^{2}}{|z_{0}|^{2}+{\cot\theta_{0}}^{2}}}(|0\rangle+\frac{\cot\theta_{0}}{z_{0}}|1\rangle) .
\nonumber
\end{eqnarray}
The parameter $z_{0}$ is yet to be determined.

The failure operators for Alice and Bob can be expressed as
\begin{eqnarray} 
A_{f}^{\dagger}A_{f}=I_{A}-|x_{0}|^{2}|r_{0}\rangle
\langle r_{0}|-|x_{1}|^{2}|r_{1}\rangle \langle r_{1}|\\
B_{f}^{\dagger}B_{f}=I_{B}-|y_{0}|^{2}|s_{0}\rangle \langle
s_{0}|-|y_{1}|^{2}|s_{1}\rangle \langle s_{1}| ,
\end{eqnarray}
where $x_{j}$, $y_{j}$, and $z_{0}$, where $j=0,1$, must be chosen so that 
these are positive operators.  The condition $A_{f}^{\dagger}A_{f}\geq 0$  
implies that
\begin{equation}
I_{A}-\frac{|x_{0}|^{2}}{1+|z_{0}|^{2}}(|0\rangle+z_{0}|1\rangle)(\langle
0|+z_{0}^{\ast} \langle
1|)-\frac{|x_{1}|^{2}}{1+|z_{0}|^{2}}(|0\rangle-z_{0}|1\rangle)(\langle
0|-z_{0}^{\ast} \langle 1|)\geq 0 ,
\end{equation}
or, in matrix form
\begin{equation}
M_{A}=\left(\begin{array}{cc}
1-\frac{|x_{0}|^{2}+|x_{1}|^{2}}{1+|z_{0}|^{2}}  & 
-\frac{z_{0}^{\ast}(|x_{0}|^{2}-|x_{1}|^{2})}{1+|z_{0}|^{2}} \\
-\frac{z_{0}(|x_{0}|^{2}-|x_{1}|^{2})}{1+|z_{0}|^{2}} & 
1-\frac{|z_{0}|^{2}(|x_{0}|^{2}+|x_{1}|^{2})}{1+|z_{0}|^{2}}
\end{array} \right) \geq 0 .
\end{equation}
This matrix will be postive if both $ {\rm Tr}M_{A}\geq 0$, which implies that
\begin{equation}
2-(|x_{0}|^{2}-|x_{1}|^{2}) \geq 0 ,
\end{equation}  
and $\det M_{A} \geq 0$, which implies
\begin{equation}
(1+|z_{0}|^{2})^{2}(1-(|x_{0}|^{2}+|x_{1}|^{2}))+4|z_{0}|^{2}|x_{0}|^{2}
|x_{1}|^{2} \geq 0 .
\end{equation}
Similar conditions are found from the requirement that $B_{f}^{\dagger}B_{f}
\geq 0$. 

Our goal is to minimize the total failure probability, $p_{f}$, which is
found by summing over all measurement results that contain a failure signal,
and is
\begin{equation}
\label{failprob}
p_{f}=\frac{1}{2}\sum_{k=0}^{1}\langle
\Psi_{k}|(A_{f}^{\dagger}A_{f}\otimes I_{B}+I_{A}\otimes
B_{f}^{\dagger}B_{f}-A_{f}^{\dagger}A_{f}\otimes B_{f}^{\dagger}B_{f})
|\Psi_{k}\rangle .
\end{equation}
We have assumed that the probability of receiving either $|\Psi_{0}\rangle$
or $|\Psi_{1}\rangle$ is the same, i.e.\ $1/2$.
We shall specialize to the case $x_{0}=x_{1}$ and $y_{0}=y_{1}$.  As we shall
see, this will still allow us to obtain the minimum achievable failure
probability.  Doing so we find that
\begin{eqnarray}
A_{f}^{\dagger}A_{f} & = & I_{A}-\frac{2|x_{0}|^{2}}{1+|z_{0}|^{2}}
(|0\rangle\langle 0|+|z_{0}|^{2}|1\rangle\langle 1|) \nonumber \\
B_{f}^{\dagger}B_{f} & = & I_{B}-\frac{2|y_{0}|^{2}|z_{0}|^{2}}{|z_{0}|^{2}
+\cot^{2}\theta_{0}}\left( |0\rangle\langle 0|+\frac{\cot^{2}\theta_{0}}
{|z_{0}|^{2}}|1\rangle\langle 1|\right)  .
\end{eqnarray}
It is clear from Eq.\ (\ref{failprob}) that the failure probability will be 
a minimum when $|x_{0}|$ and $|y_{0}|$ are as large as possible, subject 
to the constraint that the operators $A_{f}^{\dagger}A_{f}$ and 
$B_{f}^{\dagger}B_{f}$ are positive.  From the above equations, we see that
this implies that if $|z_{0}|\leq 1$, then $|x_{0}|^{2}=(1+|z_{0}|^{2})/2$
and 
\begin{equation}
\label{failA1}
A_{f}^{\dagger}A_{f}=(1-|z_{0}|^{2})|1\rangle\langle 1| ,
\end{equation} 
and if $|z_{0}|\geq 1$, then $|x_{0}|^{2}=[1+(1/|z_{0}|^{2})]/2$, and
\begin{equation}
A_{f}^{\dagger}A_{f}=\left( 1-\frac{1}{|z_{0}|^{2}}\right) 
|0\rangle\langle 0| .
\end{equation} 
We also have that if $\cot^{2}\theta_{0}\leq |z_{0}|^{2}$, then
$|y_{0}|^{2}=[1+(\cot\theta_{0}/|z_{0}|)^{2}]/2$ and 
\begin{equation}
\label{failB1}
B_{f}^{\dagger}B_{f}=\left( 1-\frac{\cot^{2}\theta_{0}}{|z_{0}|^{2}}\right) 
|1\rangle\langle 1| ,
\end{equation}
and if $\cot^{2}\theta_{0}\geq |z_{0}|^{2}$, then
$|y_{0}|^{2}=[1+(|z_{0}|/ \cot\theta_{0})^{2}]/2$ and
\begin{equation}
\label{failB2}
B_{f}^{\dagger}B_{f}=\left( 1-\frac{|z_{0}|^{2}}{\cot^{2}\theta_{0}}\right) 
|0\rangle\langle 0| .
\end{equation}

Let us consider the case when $|z_{0}|\leq 1$ and $0\leq \theta \leq \pi /4$,
which implies that Eqs.\ (\ref{failA1}) and (\ref{failB2}) apply.  We then
have that the failure probability is given by
\begin{equation}
p_{f}=1-2|z_{0}|^{2}\sin^{2}\theta_{0} ,
\end{equation}
and it is clear that this is minimized by choosing $|z_{0}|=1$.  This gives
us
\begin{equation}
p_{f}=\cos (2\theta_{0}) ,
\end{equation}
which is equal to the optimal failure probability for distinguishing the 
states $|\Psi_{0}\rangle$ and $|\Psi_{1}\rangle$. This failure probability 
is given by
\begin{equation}
1-p_{idp}=|\langle\Psi_{1}|\Psi_{0}\rangle |=\cos (2\theta_{0}) .
\end{equation}
This implies that by using this procedure, we can distinguish the states
just as well by measuring the qubits separately and comparing the results
as we can by performing a joint measurement on both of them.

Let us now summarize the results of the preceding calculations.  The states
we are distinguishing are given in Eq.\ (\ref{states}), with
$0\leq \theta \leq \pi /4$.  Alice's POVM elements are $|r_{j}\rangle
\langle r_{j}|$, for $j=0,1$, with
\begin{eqnarray}
\label{r01}
|r_{0}\rangle & = & \frac{1}{\sqrt{2}}(|0\rangle + |1\rangle ) \nonumber \\
|r_{1}\rangle & = & \frac{1}{\sqrt{2}}(|0\rangle - |1\rangle ) ,
\end{eqnarray}
and $A_{f}=0$.  This implies that Alice will only obtain the results $0$
or $1$ for her measurement, she will never receive a failure result.  In
fact, she simply performs a projective measurement.
Bob's
POVM elements are 
\begin{equation}
B_{j}^{\dagger}B_{j} = \frac{1}{2}(1+\tan^{2}\theta_{0})|s_{j}\rangle
\langle s_{j}| 
\end{equation}
for $j=0,1$, with
\begin{eqnarray}
|s_{0}\rangle & = & \sin\theta_{0}|0\rangle -\cos\theta_{0}|1\rangle  
\nonumber \\
|s_{1}\rangle & = & \sin\theta_{0}|0\rangle + \cos\theta_{0}|1\rangle ,
\end{eqnarray}
and, corresponding to the failure result,
\begin{equation}
B_{f}^{\dagger}B_{f}=(1-\tan^{2}\theta_{0})|0\rangle\langle 0|
\end{equation}.  

Examining these results, we can now see, in a simple way, how this procedure
works.  Define the single qubit states $|\psi_{j}\rangle$, for $j=0,1$ as
\begin{equation}
\label{psi01}
|\psi_{j}\rangle = \cos\theta_{0}|0\rangle +(-1)^{j}\sin\theta_{0}|1\rangle .
\end{equation}
When Alice performs her measurement, she obtains either $0$ or $1$.  If she
obtains $0$, then Bob is left with the state $|\psi_{0}\rangle$ if 
$|\Psi_{0}\rangle$ was sent, and $|\psi_{1}\rangle$ if $|\Psi_{1}\rangle$
was sent.  If she obtains $1$, then Bob is left with the state 
$|\psi_{1}\rangle$ if $|\Psi_{0}\rangle$ was sent, and $|\psi_{0}\rangle$ 
if $|\Psi_{1}\rangle$ was sent.  In either case, Bob is faced with 
discriminating between the non-orthogonal states $|\psi_{0}\rangle$ and
$|\psi_{1}\rangle$.  He then applies the optimal POVM to distinguish 
between these states, and if he succeeds, he knows which of the two states
he has.  What he does not know, is which of his single-qubit states
corresponds to $|\Psi_{0}\rangle$, and which to $|\Psi_{1}\rangle$.  It is
this bit of information that the result of Alice's measurement contains.
Only by combining the results of their measurements can Alice and Bob
deduce which state was sent.

The analysis in the preceding paragraph immediately allows us to see that 
there is another solution to the problem of finding a POVM in which one
of the parties can receive a failure signal, and that is the one in which
the roles of Alice and Bob are interchanged.  In that case, Bob makes a
projective measurement, and Alice makes a measurement whose results are
described by a three-outcome POVM.

It was noted by Virmani, \emph{et al}. \cite{virmani}, that for 
any two two-qubit states
with the same Schmidt basis, which they called Schmidt correlated, it is
possible for Alice to transfer all of the information about the state
to Bob by making a measurement in the basis $\{ |r_{0}\rangle ,|r_{1}
\rangle\}$ and telling Bob the result of her measurement.  In general 
Bob's measurement will depend on the results of Alice's.  What we have
seen in this section is that for special choices of the two states,
Alice and Bob always make the same measurement, which means they can 
make the measurement as soon as they receive the particles.  They, each,
then, posses a classical bit, and by comparing these bits they can tell
which state they were sent.  
 
\section{Optical realization}
We now want to show how this measurement can be realized optically.  The
states $|\Psi_{0}\rangle$ and $|\Psi_{1}\rangle$ are two-photon states
with the information encoded in the polarization of the photons.  We
suppose that $|0\rangle$ corresponds to horizontal polarization and
$|1\rangle$ to vertical.  Alice's measurement is then straightforward;
she sends her photon through a polarization beam splitter.  A horizontally
polarized photon incident on this device will continue in a straight line
while a vertically polarized photon will be deflected by ninety degrees.
Alice orients her polarization beam splitter so that a photon in the 
polarization state $(|0\rangle +|1\rangle )/\sqrt{2}$ is transmitted
and one in the state $(|0\rangle -|1\rangle )/\sqrt{2}$ is deflected.  She
has detectors in both paths, and she simply observes which one clicks.

\begin{figure}
\epsfig{file=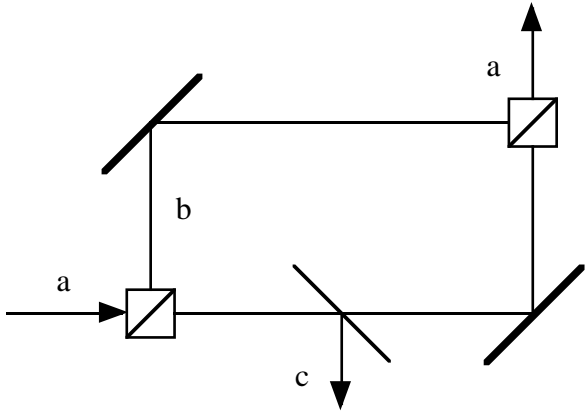}
\caption{Interferometer for realizing three-outcome POVM.  The photon 
enters in mode $a$ and encounters a polarization beam splitter.  Its
vertically polarized part is refelected into mode $b$, while its
horizontally polarized part is transmitted in mode $a$.  In its passage
through the device, the photon encounters a polarization insensitive
beam splitter and one more polarization beam spliter.  If the photon
emerges in mode $c$, the measurement has failed, and if it emerges in
mode $a$, it has succeeded.}
\end{figure}

Bob's measurement is more complicated, but it has been worked out by
Huttner, \emph{et al}. \cite{huttner}.  They presented two implementations
of the POVM, one in which the failure signal can be detected explicitly and
one in which it cannot, and demonstrated the second experimentally.  We
shall describe their first scheme.  It makes use of two polarization
beam splitters, and one standard, polarization-insensitive beam splitter,
and is depicted in Fig.\ 1.  The input state, which is either 
$|\psi_{0}\rangle$ or $|\psi_{1}\rangle$, is sent into the first polarization
beam splitter in mode $a$.  The vertically polarized part of the state is
deflected into mode $b$, while the horizontally polarized part continues
in mode $a$.  If the input state is given by $|\phi_{in}\rangle_{a}
=\alpha |0\rangle_{a}+\beta |1\rangle_{a}$, where the subscripts on the
states denote the mode, we have that just after the first polarization
beam splitter
\begin{equation}
|\phi_{in}\rangle_{A}\rightarrow \alpha |0\rangle_{a}+\beta |1\rangle_{b}.
\end{equation}
The beam splitter transmits a photon with transmissivity $t$ and reflects it
with reflectivity $r$.  This implies that after passing through the
beam splitter the state $|0\rangle_{a}$ becomes $t|0\rangle_{a}+
r|0\rangle_{c}$.  Finally, after the second polarization beam splitter, the
output state, $|\phi_{out}\rangle$ is
\begin{equation}
|\phi_{out}\rangle = \alpha t|0\rangle_{a}+\beta |1\rangle_{a}+\alpha r
|0\rangle_{c} .
\end{equation}
Choosing $t=\tan\theta_{0}$, we have that if the input state is 
$|\psi_{0}\rangle$, then
\begin{equation}
|\phi_{out}\rangle = \sin\theta_{0} (|0\rangle_{a}+|1\rangle_{a})
+\sqrt{\cos 2\theta_{0}} |0\rangle_{c} ,
\end{equation}
and if the input state is $|\psi_{1}\rangle$, then
\begin{equation}
|\phi_{out}\rangle = \sin\theta_{0} (|0\rangle_{a}-|1\rangle_{a})
+\sqrt{\cos 2\theta_{0}} |0\rangle_{c} .
\end{equation}
Note that the parts of the two output states in the $a$ mode have orthogonal
polarizations, and can be distinguished by orienting a third polarzation
beam splitter so that $(|0\rangle_{a} +|1\rangle_{a} )/\sqrt{2}$ is 
transmitted and $(|0\rangle_{a} -|1\rangle_{a} )/\sqrt{2}$ is deflected.
If the photon is detected in mode $c$, the procedure has failed.  Note
that both Alice's and Bob's measurements can be realized using only
linear optics.

\section{More than two parties}
It is relatively easy to generalize the procedure in section 3 to divide the
information about which of two states was sent among any number of parties.
We shall show how to do this for both qubits and for qutrits.

Let us start with two $N$-qubit states
\begin{eqnarray}
|\Psi_{0}\rangle & = & \cos\theta_{0} |00\ldots 0\rangle +\sin\theta_{0} 
|11\dots 1\rangle \nonumber \\
|\Psi_{1}\rangle & = & \cos\theta_{0} |00\ldots 0\rangle -\sin\theta_{0} 
|11\dots 1\rangle ,
\end{eqnarray}
where $0\leq \theta_{0} \leq \pi /4$.  Each of the qubits is sent to one of
$N$ parties, $A_{1}, \ldots A_{N}$.  Each of the parties, $A_{1}$ through
$A_{N-1}$ measures their qubit in the $\{ r_{0},r_{1}\}$ basis (see Eq.\ 
(\ref{r01})), and $A_{N}$ performs the unambiguous-state discrimination
procedure for the states $|\psi_{0}\rangle$ and $|\psi_{1}\rangle$ 
(see Eq.\ (\ref{psi01})).
If parties $A_{1}$ through $A_{N-1}$ obtained $n_{0}$ results of 
$|r_{0}\rangle$ and $n_{1}$ results of $|r_{1}\rangle$, then the states
that $A_{N}$ is distinguishing between are
\begin{eqnarray}
|\psi_{0N}\rangle & = & \cos\theta_{0} |0\rangle +(-1)^{n_{1}}\sin\theta_{0} 
|1\rangle \nonumber \\
|\psi_{1N}\rangle & = & \cos\theta_{0} |0\rangle -(-1)^{n_{1}}\sin\theta_{0} 
|1\rangle ,
\end{eqnarray}
i.e.\ $A_{N}$'s qubit will be in the state $|\psi_{0N}\rangle$ 
if the state $|\Psi_{0}\rangle$ was sent and $|\psi_{1N}\rangle$ if 
the state $|\Psi_{1}\rangle$ was sent.  In order to
ascertain which of the two $N$-qubit states was sent, all of the parties will
have to combine their information.  If the measurement made by $A_{N}$
succeeds, then she will have obtained either 
$|\psi_{0}\rangle$ or $|\psi_{1}\rangle$,
but she will not, without knowing the measurement results of all of the
other parties, know which of these results corresponds to $|\Psi_{0}\rangle$
and which corresponds to $|\Psi_{1}\rangle$.

The procedure can be generalized to particles with more than two internal
states, and to demonstrate this we shall consider the case of qutrits.
Consider the three $N$-qutrit states
\begin{eqnarray}
|\Psi_{0}\rangle & = & c_{0}|0\ldots 0\rangle +c_{1}|1\ldots 1\rangle 
+c_{2}|2\dots 2\rangle \nonumber \\
|\Psi_{1}\rangle & = & c_{0}|0\ldots 0\rangle +c_{1}\omega |1\ldots 1\rangle 
+c_{2}\omega^{\ast}|2\dots 2\rangle \nonumber \\
|\Psi_{2}\rangle & = & c_{0}|0\dots 0\rangle +c_{1}\omega^{\ast} 
|1\ldots 1\rangle +c_{2}\omega |2\ldots 2\rangle ,
\end{eqnarray}
where $\omega =\exp (2\pi i/3)$.  Define the single qutrit orthonormal
basis
\begin{eqnarray}
|\eta_{0}\rangle & = & \frac{1}{\sqrt{3}}(|0\rangle + |1\rangle + |2\rangle )
\nonumber \\
|\eta_{1}\rangle & = & \frac{1}{\sqrt{3}}(|0\rangle + \omega |1\rangle + 
\omega^{\ast}|2\rangle )\nonumber \\
|\eta_{2}\rangle & = & \frac{1}{\sqrt{3}}(|0\rangle + \omega^{\ast}|1\rangle 
+ \omega |2\rangle ) .
\end{eqnarray}
Each of the $N$ qutrits is sent to one of the parties $A_{1}$, \ldots $A_{N}$. 
Now, parties $A_{1}$ through $A_{N-1}$ perform projective measurements in
the basis $\{|\eta_{0}\rangle ,|\eta_{1}\rangle , |\eta_{2}\rangle \}$, 
and suppose that $m_{j}$ of them find their qutrit in the 
state $|\eta_{j}\rangle $, $j=0,1,2$.  The party $A_{N}$
performs the optimal POVM to unambiguously distinguish the states
\cite{terno,sun}
\begin{eqnarray}
|\psi_{0}\rangle & = & c_{0}|0\rangle +c_{1}|1\rangle +c_{2}|2\rangle 
\nonumber \\
|\psi_{1}\rangle & = & c_{0}|0\rangle +c_{1}\omega |1\rangle 
+c_{2}\omega^{\ast}|2\rangle \nonumber \\
|\psi_{2}\rangle & = & c_{0}|0\rangle +c_{1}\omega^{\ast} |1\rangle 
+c_{2}\omega |2\rangle .
\end{eqnarray}
After the parties $A_{1}$ through $A_{N-1}$ have performed their measurements,
the qutrit belonging to $A_{N}$ is in one of the three states
\begin{eqnarray}
|\psi_{0N}\rangle & = & c_{0}|0\rangle +c_{1}\omega^{(m_{2}-m_{1})}|1\rangle 
+c_{2}\omega^{-(m_{2}-m_{1})} |2\rangle 
\nonumber \\
|\psi_{1N}\rangle & = & c_{0}|0\rangle +c_{1}\omega^{(m_{2}-m_{1}+1)}|1\rangle 
+c_{2}\omega^{-(m_{2}-m_{1}+1)} |2\rangle \nonumber \\
|\psi_{2N}\rangle & = & c_{0}|0\rangle +c_{1}\omega^{(m_{2}-m_{1}-1)}|1\rangle 
+c_{2}\omega^{-(m_{2}-m_{1}-1)} |2\rangle .
\end{eqnarray}
The qutrit is in the state $\psi_{jN}$ if the original $N$-qutrit state 
was $\Psi_{j}$, for $j=0,1,2$.

If the measurement made by $A_{N}$ succeeds, she will have found her qutrit
in one of the states $\psi_{j}$, $j=0,1,2$.  She will not know to which of
the original $N$-qutrit states it corresponds, however, without knowing
the measurement results of all of the other parties.  In particular, we
have the correspondence
\begin{equation}
\Psi_{j}\leftrightarrow \psi_{[j+m_{2}-m_{1}\ {\rm mod}3]} .
\end{equation}
Therfore, all of the parties must combine their information in order to
determine which of the three $N$-qutrit states was originally sent.

Note that in both the case of $N$ qubits and $N$ qutrits, only one party 
will receive a failure signal if the measurement fails.  In addition,
the probability of failure is the best possible, i.e.\ 
it is the same as it would be if all of the qubits or qutrits were 
measured together.  Consequently, we have not lost anything by measuring
the particles separately.

\section{Conclusion}
We have shown that it is possible to distinguish two non-orthogonal 
two-qubit states by local measurements and classical communication, making 
no errors and with one of the parties receiving a failure signal if the 
procedure fails.  Both of the parties make fixed measurements, it is not
the case that the measurement made by one party depends on the result
obtained by the other.  If the procedure succeeds, each party obtains
either a $0$ or a $1$, and gains no information about the state from
their individual results.  However, on combining their results, the parties
can identify the state.

This procedure should be useful as a basis for quantum secret sharing.  It 
provides security in the same way as does the B92 protocol for quantum key 
distribution \cite{bennett92}.  An
eavesdropper, Eve, who intercepts the two-qubit state cannot indentify it 
with certainty.  The best she can do is to apply the two-state unambiguous
state discrimination procedure, which will sometimes fail.  When it does,
she does not know which state to send on to Alice and Bob, and will,
consequently, introduce errors, e.g.\ Alice and Bob will have detected
$|\Psi_{0}\rangle$ when $|\Psi_{1}\rangle$ was sent.  These errors can
be detected if Alice and Bob publicly compare a subset of their measurements
with information provided by the person who sent the states. 

There is also some protection against cheating.  If Alice cheats by obtaining
both qubits, then the best she can do is to apply two-state unambiguous
state discrimination to them.  Her measurement will sometimes fail, and
then she has a problem.  She must send a qubit to Bob, but there is no state
for this qubit that will make Bob's measurement fail with certainty.  That 
means that Bob will sometimes obtain incorrect results, i.e.\ when he and
Alice combine their results, they will find that the state they detected was
not the one that was sent.  Therefore, cheating by Alice will introduce 
errors.

If Bob has obtained both particles, then he also can apply two-state 
unambiguous state discrimination to the two-qubit state.  If his 
measurement succeeds, he can just send a qubit in the appropriate state
to Alice, and if it fails, he can simply state that it failed.  That
means that cheating by Bob cannot be detected.  However, a 
modification of the protocol will solve this problem.  When the 
two-qubit state is sent, the person sending the state can announce over
a public channel, which of the parties is to make the projective measurement
and which is to make the three-outcome POVM.  This means that part of the
time, Bob will be assigned to make the projective measurement, and then 
his cheating will be detected.  He can, however, not cheat if he is
assigned to make the projective measurement, and in that case he will
gain partial information about the key and not be detected.  One way to
address this problem is to combine several received 
bits into a block, the parity of which is a single key bit.  In order for
Bob to ascertain the key bit, he would have to know all of the received bits
in the block, but the probability that he would can be made very low by
choosing the block size sufficiently large.  

Secret sharing, then, provides one application of the state discrimination 
procedures discussed in this paper.  Whether there are others is a subject
for future work. 

\section*{Acknowledgments}
This research was supported by the National Science Foundation under
grant number PHY 0139692.

\bibliographystyle{unsrt}

\end{document}